\documentstyle[eqsecnum,floats,preprint,epsf,aps]{revtex}
\input{epsf}

\tighten

\begin{document}

\renewcommand{\arraystretch}{1.5}
\newcommand{\be}{\begin{equation}}
\newcommand{\ee}{\end{equation}}
\newcommand{\bea}{\begin{eqnarray}}
\newcommand{\eea}{\end{eqnarray}}
\newcommand{\la}{\leftarrow}
\newcommand{\ra}{\rightarrow}
\newcommand{\lr}{\leftrightarrow}
\newcommand{\La}{\Leftarrow}
\newcommand{\Ra}{\Rightarrow}
\newcommand{\Lr}{\Leftrightarrow}
\def\Tr{\mathop{\rm Tr}\nolimits}
\def\mapright#1{\smash{\mathop{\longrightarrow}\limits^{#1}}}
\def\mapdown#1{\big\downarrow \rlap{$\vcenter
  {\hbox{$\scriptstyle#1$}}$}}
\def\Z{{\bf Z}}
\def\R{{\bf R}}
\def\M{{\cal M}}
\def\L{{\cal L}}
\def\c{c_\chi}
\def\s{s_\chi}
\def\xh{\hat x}
\def\yh{\hat y}
\def\zh{\hat z}
\def\vt{\widetilde{v}}
\def\wt{\widetilde{w}}
\def\n#1{{\hat{n}_{#1}}}
\def\c#1{\cos{#1}}
\def\s#1{\sin{#1}}
\def\cs#1{\cos^2{#1}}
\def\ss#1{\sin^2{#1}}
\newcommand{\PSbox}[3]{\mbox{\rule{0in}{#3}\includegraphics{#1}\hspace{#2}}}

\title{Black Holes and Instabilities of Negative Tension Branes}

\author{Donald Marolf\footnote{marolf@physics.syr.edu} and
Mark Trodden\footnote{trodden@physics.syr.edu}}

\address{Department of Physics \\
Syracuse University \\
Syracuse, NY 13244-1130, USA. \\
}

\maketitle

\begin{abstract}
We consider the collision in 2+1 dimensions of a black hole and
a negative tension brane on an orbifold.  Because there is no
gravitational radiation in 2+1 dimensions, the horizon area shrinks
when part of the brane falls through.   This provides a
potential violation of the generalized second law of thermodynamics.
However, tracing the details of the dynamical evolution one finds
that it does not proceed from equilibrium configuration to equilibrium
configuration.  Instead, a catastrophic
space-time singularity develops similar to the `big crunch' of $\Omega >1$
FRW space-times.
In the context of classical general relativity,
our result demonstrates a new instability of constructions
with negative tension branes.
\end{abstract}

\setcounter{page}{0}
\thispagestyle{empty}

\vfill

\noindent SU-GP-01/2-2 \hfill hep-th/01?????

\hfill Typeset in REV\TeX

\eject

\vfill

\eject

\baselineskip 20pt plus 2pt minus 2pt

\section{Introduction}
\label{sec:introduction}
The idea that our $3$ familiar spatial dimensions may exist as a submanifold of
a higher dimensional space-time has opened a number of new avenues for string
theory, particle physics and cosmology
\cite{Rubakov:1983bb}-\cite{Corley:2001rt}. Common
to most of the recent incarnations is the condition that all interactions other than gravity be confined
to our submanifold, or {\it brane}. Since departures from $3+1$ dimensional gravity are relatively difficult
to constrain compared to those of
other forces, this permits significant freedom to modify the gravitational
interactions in the extra-dimensional space. It is in this modification of gravity and of the
extra-dimensions themselves that recent approaches differ from one another.

In many of the proposed models, the hierarchy problem is recast by bringing the fundamental mass scale of
physics down to the weak scale. The large Planck mass observed on our $3$-brane is then a derived
quantity, the size of which arises from the relatively large volume of the extra-dimensional manifold. A
striking consequence of this is that one expects quantum gravity or stringy effects to manifest themselves
close to the electroweak scale. At first sight, this opens up many interesting possibilities for
testing the idea of extra dimensions \cite{Giudice:1999ck}-\cite{Davoudiasl:2000my}. However, it is typically possible to construct these models
consistently in such a way that they evade expected experimental tests.

One specific hurdle that extra-dimensional theories must clear is that the brane-bulk system
should be a consistent, stable solution to Einstein gravity. In this paper we consider this issue for those
constructions that include negative-tension branes. While the notion of
negative energies is typically problematic, perturbative dynamical
objections can be overcome by placing the offending brane
at an orientifold plane.

In this paper we address two other issues associated with negative-tension
branes, one of which is specific to those at orbifold fixed planes.
The first has to do with their consequences for the
generalized second law of thermodynamics, which
states that the total entropy in matter and black holes does
not decrease. In Einstein gravity this entropy can be written as
\begin{equation}\label{gensecondlaw}
  S_{\rm TOT} \equiv S_{\rm matter} + \frac{1}{4}A \ ,
\end{equation}
where A is the sum of the event horizon areas of all black holes, and we are using Planck units
($\hbar=c=k=G=1$). In particular, any process which leaves $S_{\rm matter}$ fixed and decreases $A$ leads
to a violation of the generalized second law.

We begin with the simple observation that the area of a black hole event horizon increases when positive energy
crosses the horizon, and decreases when negative energy crosses the horizon. A natural question then
arises when one has a space-time with a negative tension brane: what happens if a black hole,
initially far away from
the brane, falls towards the brane and captures some of the brane within its horizon? One might expect that
the generalized second law could be violated in this way, since the part of
the brane that is swallowed by
the black hole carries negative energy across the horizon.

In general, of course, whether the horizon actually shrinks will depend
on what other matter or gravitational radiation is falling into the black
hole.  For this reason we consider a
lower-dimensional system; a negative
tension $1$-brane in a $2+1$ dimensional space-time.
The absence of gravitational radiation in $2+1$
dimensions will force the conclusion that the horizon shrinks
when the black hole encounters the negative tension brane.
This is so whether or not the brane sits at an orbifold.

This would provide a clear violation of the second law if
the above collision connects two equilibrium configurations.
Whether or not this is so turns out to be an interesting question that
our low dimensional context allows us to explore in detail.
In $2+1$ dimensions, the only black holes are the so called
BTZ black holes of Ba\~{n}ados, Teitelboim and Zanelli \cite{Banados:1992wn},
which arise only for negative cosmological constant.
These black hole space-times can be constructed as a quotient of
AdS space itself.  Together with
the lack of local gravitational degrees of freedom in 2+1 dimensions,
this fact allows us to construct the general solution describing the
collision of a black hole and a brane.   Interestingly, for branes at
$Z_2$ orbifolds the results depend
dramatically on the sign of the brane tension.

We investigate a 2-parameter family of solutions with negative tension
branes in detail.  This leads to our second issue.  We find that
the endpoint is not in fact an equilibrium configuration but instead
is a space-like singularity similar to the `big crunch' of
FRW models with $\Omega > 1$ (and no cosmological constant).
Continuity implies that a similar result occurs
on an open subset of the full space of solutions, and it seems likely that
the result is generic.  This is naturally interpreted as a non-linear
dynamical instability of gravitating negative tension branes at orbifolds.

The paper is organized as follows. In the next section we provide a brief calculation of how the black hole
horizon shrinks in our model.
In section~\ref{sec:space-time} we then describe how to construct
initial data for the brane - black hole system in AdS$_{2+1}$ and investigate
the dynamics of the system in detail. This dynamics is qualitatively
different in various regions of parameter space.
We conclude in section~\ref{sec:conclusions}
and offer some interpretations of our results.

\section{The Shrinking Horizon}
\label{sec:raychauduri}
Our purpose in this section is to give a description of how
a horizon evolves as it falls
towards the negative tension brane.
A black hole horizon is a closed null hypersurface.
If we consider a congruence of null geodesics with
tangent vector $N^a$ orthogonal (and therefore tangent) to this hypersurface,
then the divergence $\theta\equiv \nabla_a N^a$,
describing the expansion or contraction of the congruence is described by the Raychaudhuri equation
\begin{equation}\label{raychaudhuri}
  \frac{d\theta}{d\lambda}=-\frac{1}{2}\theta^2 - \sigma_{ab}\sigma^{ab} + \omega_{ab}\omega^{ab}
  -R_{ab}N^a N^b \ ,
\end{equation}
where $\sigma_{ab}$ is the shear of the congruence,
$\omega_{ab}$ is its twist, and $\lambda$ is the affine
parameter.
(For a detailed description of this equation see, for example, \cite{Wald}.)
The metric used to raise and lower the indices $a,b$ is the induced metric
on the null surface, which has signature (0,+,+).
As a result, the twist term is positive definite.
The shear term is related to the Weyl tensor, and
hence to the existence of gravitational radiation --
all of which vanish in $2+1$ dimensions.  Therefore, from now on
we set $\sigma_{ab} \equiv 0$.

Now, an infinitesimal area element $dA$ of the event horizon of
the black hole is related to $\theta$ via
\begin{equation}\label{area}
  dA=dA_0 \exp\left(\int \theta(\lambda) \, d\lambda\right) \ ,
\end{equation}
and therefore the Raychaudhuri equation can be used to describe the evolution of the area as
\begin{equation}\label{areaevolution}
  \frac{d^2}{d\lambda^2}(\sqrt{A}) = \left(
\omega_{ab} \omega^{ab} - \frac{1}{2} R_{ab}N^a N^b \right) \sqrt{A} \ .
\end{equation}
Using the Einstein equation, and recalling that $N^a$ is null, we may rewrite this as
\begin{equation}\label{areaevolution2}
  \frac{d^2}{d\lambda^2}(\sqrt{A}) = \left( \omega_{ab} \omega^{ab} -
4\pi T_{ab}N^a N^b \right) \sqrt{A} \ ,
\end{equation}
where $T_{ab}$ is the energy momentum tensor.
Since the brane has negative tension, so long as no other matter is present we have
\begin{equation}\label{nullenergy}
  T_{ab}N^a N^b <0 \ ,
\end{equation}
for null $N^a$ not tangent to brane
(i.e. the null energy condition is violated).

If the system were to reach equilibrium and a final horizon
could be identified, the final expansion $\theta$ would vanish.
It then follows from (\ref{areaevolution2}) and (\ref{nullenergy})
that $\theta$ would have been negative during the collision itself.
In particular, no caustics can form on this surface during the collision
since immediately after a caustic the expansion of
the corresponding null generator must be positive.
Thus, the null surface ${\cal S}^{\rm final}$
that eventually becomes the final horizon can be extended back to before
the collision occurred, and we see that the area of  ${\cal S}^{\rm final}$
was larger before the collision than it is afterward.

Similarly, suppose that the system began in equilibrium and that an initial
horizon could be identified.  Any outward directed
null congruence outside this initial
horizon has positive expansion at this stage and we see that this
will continue during the collision.  Thus,
the null surface ${\cal S}^{\rm final}$
associated with the final horizon must begin inside the initial horizon.
As a result, the initial area of ${\cal S}^{\rm final}$ must be smaller
than that of the initial horizon.  Since the area of ${\cal S}^{\rm final}$
decreases during the collision, the final area must be even smaller.
We therefore arrive at a violation of the generalized second law if
the system evolves from one equilibrium to another.

\section{The Brane-BTZ Black Hole System}
\label{sec:space-time}

The discussion in the previous section describes the process of a
black hole colliding with a negative tension brane and traces the evolution
of null congruences.
On the surface, this would appear to yield a violation
of the second law of thermodynamics.  However, in order to identify
particular null congruences as initial and final `horizons,' the evolution
must connect two equilibrium configurations.

Whether or not this is
the case in our 2+1 setting is explored below through detailed investigation of
the corresponding solutions.
We require that our solutions contain a flat brane at an orbifold fixed plane and
a BTZ black hole, but the space-time should be otherwise empty.
Because $2+1$ dimensional gravity contains no local dynamics,
the metric will be locally AdS everywhere except at the orbifold singularity.

At this point it is convenient to recall that the BTZ
space-time is itself a quotient of AdS space. Similarly, recall that a space-time
(the `brane-only space-time') describing the
negative tension brane by itself consists of two copies of AdS patched together along the brane with
$Z_2$ orbifold boundary conditions.
Now, of particular interest are cases in which the black hole and brane are
initially separated.  By this we mean that at least near some Cauchy surface $t=0$
there should be a region ${\cal R}_{\rm BTZ}$ of space-time enclosing our black hole in which
the metric takes the standard BTZ form. Similarly, there should exist a
region ${\cal R}_{\rm brane}$, near the brane.
Since the space-time is to contain no further topological
complications we may take the union ${\cal R}_{\rm brane} \cup {\cal R}_{\rm BTZ}$
to contain a Cauchy surface for the space-time and the intersection
${\cal R}_{\rm brane} \cap {\cal R}_{\rm BTZ}$ to be contractible.

Let us now count the number of parameters needed to describe our space-time.
The tension of the brane is fixed by the cosmological constant, so there
are no parameters associated with the brane region ${\cal R}_{\rm brane}$ itself.  The
BTZ region ${\cal R}_{\rm BTZ}$ is determined by two free parameters corresponding
to the mass and spin of the BTZ black hole.
Since the BTZ space-time and the brane-only
space-time are by themselves locally AdS, the intersection
${\cal R}_{\rm brane} \cap {\cal R}_{\rm BTZ}$ is associated with an element of
the AdS isometry group SO(2,2) which serves to patch the two regions together.

Now consider a `fundamental region' ${\cal F}_{\rm BTZ}$ of AdS space
that covers the BTZ space-time exactly once under the BTZ quotient map. Further,
in the brane-only space-time the AdS region
on one side of the brane is related to that on the other side through the
orbifolding procedure. Hence, we refer to the region on one side of the brane as
the `fundamental region' ${\cal F}_{\rm brane} \subset$ AdS.
We may therefore write ${\cal R}_{\rm BTZ} \subset {\cal F}_{\rm BTZ}$
and ${\cal R}_{\rm brane} \subset {\cal F}_{\rm brane}$, provided
that we keep in mind the appropriate identifications.

The choice of ${\cal F}_{\rm BTZ}$ and ${\cal F}_{\rm brane}$
is, of course, unique only up to an element of the AdS symmetry group SO(2,2).
However, we may use this symmetry to bring one of the fundamental regions
into a standard configuration.
Thus, only that element of SO(2,2) describing the relative
configuration of ${\cal F}_{\rm BTZ}$ and
${\cal F}_{\rm brane}$ in AdS is of interest.  This is precisely
the freedom associated with the overlap of ${\cal R}_{\rm BTZ}$ and
${\cal R}_{\rm brane}$ in our space-time.

Consider then the set of all space-times constructed by
specifying BTZ and brane
fundamental regions ${\cal F}_{\rm BTZ}$ and ${\cal F}_{\rm brane}$ in AdS$_{2+1}$, taking
the intersection ${\cal F}_{\rm BTZ} \cap {\cal F}_{\rm brane}$ and
making appropriate identifications.   We have shown that all of the space-times
we seek will lie in this set.  However, some space-times constructed in this
way may fail to be smooth or may fail to contain separated branes and black holes.
Thus, we may say that the identification of ${\cal F}_{\rm BTZ}$ and
${\cal F}_{\rm brane}$ is necessary but not sufficient for building the space-times
of interest.

It will be enough for us to analyze the most straightforward case in which
$t=0$ is a moment of time symmetry.  From now on we will assume that this
condition holds.  As a result, we now restrict consideration to spinless
BTZ black holes.

\subsection{Initial Data}

\label{iD}

Much of the physics of this problem can be understood by studying the
initial data associated with the moment of time symmetry.   If $t$ is a global
time coordinate on AdS$_{2+1}$, we may take this moment to occur in the slice
$t=0$.  Here it is convenient to consider AdS$_{2+1}$ as a solid cylinder so
that the $t=0$ surface is a disk.  Given the condition that ${\cal F}_{\rm BTZ}$
and ${\cal F}_{\rm brane}$ respect this time symmetry,
the identifications of
AdS used to make the BTZ or brane space-times preserve the surface $t=0$.
As a result, we may speak about the fundamental domains ${\cal F}^0_{\rm BTZ}$
and ${\cal F}^0_{\rm brane}$ associated with the corresponding
quotient of the $t=0$ slice.

The space of time symmetric initial data is labelled by four parameters.
To see this, recall that the black hole spin has been set to zero and
that the requirement that the brane and black hole both be present at $t=0$
forces the SO(2,2) parameter used to patch together
${\cal R}_{\rm BTZ}$ and ${\cal R}_{\rm brane}$
to take values in $SO(2, 1)$, the symmetry group of the $t=0$ surface itself.
However, it is sufficient for our purposes to work with the two parameter
sub-family in which we require the space-time to have an additional
$Z_2$ symmetry as described below.  This requirement breaks the $SO(2,1)$
down to just the one-dimensional subgroup associated with one of the boost
generators, and we may think of the
associated free parameter as the location of the brane in
Poincar\'e coordinates.

For definiteness we will work in
global coordinates $(t,\rho, \phi)$ in which the AdS line element takes the form
\begin{equation}
\label{sausagemetric}
ds^2 = \frac{4\ell^4}{(\ell^2 -\rho^2)^2} \left[ -
\left( \frac{\ell^2 + \rho^2}{2 \ell^2} \right)^2 dt^2 + d\rho^2 + \rho^2 d \phi^2 \right] \ ,
\end{equation}
with $t\in [-\pi/2,\pi/2]$, $\rho \in [0,1]$ and $\phi \in [-\pi,\pi]$.
These are the coordinates referred to as `sausage coordinates'
in \cite{Swed}, which is a useful reference for visualizing the BTZ
fundamental region. The extra $Z_2$ symmetry we impose corresponds
to invariance under $\phi \rightarrow -\phi$.

\begin{figure}[ht]
\centerline{ \epsfbox{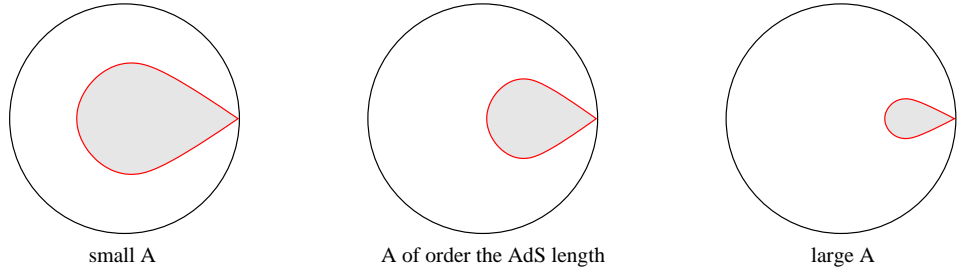}}
\caption{The brane fundamental region at $t=0$ for three
values of $A$.}
\label{figure1}
\end{figure}

\begin{figure}[ht]
\centerline{\epsfbox{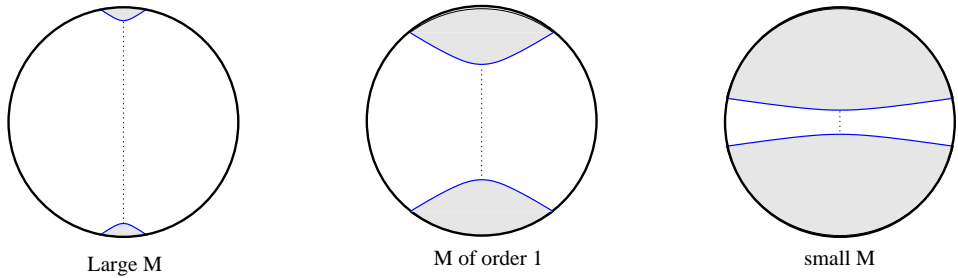}}
\caption{The BTZ fundamental region at $t=0$ for three
values of $M$.}
\label{figure2}
\end{figure}

Using these coordinates,  ${\cal F}^0_{\rm brane}$ and ${\cal
F}^0_{\rm BTZ}$ are drawn in figure~\ref{figure1} and
figure~\ref{figure2} respectively.  The fundamental domains are
the unshaded regions. Each figure shows three sample cases
illustrating what happens when we change the free parameter
associated with the respective fundamental region. In
figure~\ref{figure1}, the parameter (which we call $A$) may be
thought of as the location of the brane in Poincar\'e coordinates.
The three cases shown there are of course related by the action of
an AdS isometry, but this will shortly be broken by the presence
of the black hole. In figure~\ref{figure2}, the parameter is the
mass $M$ of the BTZ black hole, which controls the `size' of the
fundamental region as shown.  The dotted line in
figure~\ref{figure2} is the intersection of the black hole horizon
with the $t=0$ surface.  This intersection is the bifurcation
surface of the horizon.

It remains to put these domains
together in a consistent way.  We are interested in
configurations in which the black hole and brane are separated at $t=0$,
meaning that the brane lies entirely on one side of the black hole horizon.
It is clear that for any black hole mass $M$ one can choose a large enough
value of $A$ so that the objects are indeed separated.  Note that
rotating the disks in figure~\ref{figure2} by the action $\phi\rightarrow \phi +\pi/2$ would also yield
a configuration
compatible with our symmetries, but that it would then be impossible to separate
the brane and black hole\footnote{One can also show that
such configurations are singular at $t=0$.}.

It is interesting to note that positive tension branes behave
very differently.  We may represent a positive tension brane as in figure~\ref{figure1}
above, but with the shaded region now representing the fundamental
region.  A glance at figures 1 and 2 shows that it is now impossible to
separate brane and black hole.  This is clearly a reflection of the well-known
statement that there is only a `small' region of AdS that is far from a positive
tension brane on a $Z_2$ orbifold.

\subsection{Dynamics and Collisions}

\begin{figure}[ht]
\centerline{ \epsfbox{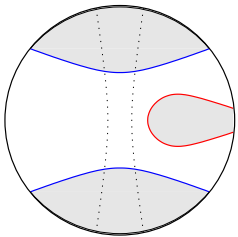}}
\caption{A typical configuration of
${\cal F}_{\rm BTZ}$ and ${\cal F}_{\rm brane}$ at $t>0$.}
\label{figure3}
\end{figure}

In section \ref{iD} we
established the existence of initial data representing
a separated, momentarily static configuration containing a black hole and
a negative tension brane.  We now wish to study the time evolution of this
solution.  Certain qualitative features are readily deduced from the well-known
properties of black holes and branes with respect to the global
coordinates $(t,\rho, \phi)$ on AdS$_{2+1}$.  For example, as $t$
increases the black hole horizon expands outward to both right and left, while
the boundaries of the BTZ fundamental region move toward each other (see
e.g. \cite{Swed}).  On the other hand, the brane expands, with the ends separating
and moving along the boundary at the speed of light.  A typical configuration
of both black hole and brane for $t >0$ is shown in figure~\ref{figure3}.

The expansion of the brane and the contraction of the BTZ fundamental region
clearly results in contact between the boundaries of ${\cal F}_{\rm BTZ}$
and ${\cal F}_{\rm brane}$.
The points of contact will be conical singularities
in the resulting space-time.
If we simply extrapolate the behavior of these
boundaries it would appear that the entire space-time outside the black hole
would quickly disappear
into such a singularity.  However, once the singularity forms
it is no longer clear that
we can follow the evolution using classical General Relativity.
It is therefore important to discover whether events along this potential
singularity are causally connected, so that we may determine whether
modifications to General Relativity at one such point could affect the
formation of the rest of the singularity.

This turns out to be simplest to analyze in Poincar\'e coordinates, which
we label $\gamma$, $\beta,$ and $z$.  The relevant metric is most easily obtained by making
use of `embedding' coordinates
as an intermediate step. Recall that AdS$_{2+1}$ may be
described as the surface
\begin{equation}
\label{adsembeddedsurface}
T^2 + U^2 -X^2 -Y^2 = \ell^2 \ ,
\end{equation}
embedded in $2+2$ dimensional Minkowski space, with line element
\begin{equation}
ds^2 = -dT^2 -dU^2 +dX^2 +dY^2 \ .
\end{equation}
In (\ref{adsembeddedsurface}), $\ell$ is the AdS length scale.
The embedding coordinates $(T,U,X,Y)$ are related to our global coordinates via
\begin{eqnarray}
X = - \frac{2 \ell^2  \rho}{\ell^2 - \rho^2} \cos (\phi) \ ,\ \ \
&\ &\ \ \ T =  \ell\left[\frac{\ell^2 + \rho^2}{\ell^2 -
\rho^2}\right] \sin \left(\frac{t}{\ell}\right) \cr Y = \frac{2
\ell^2 \rho}{\ell^2 - \rho^2} \sin (\phi) \ ,\ \ \ &\ &\ \ \ U =
\ell\left[\frac{\ell^2 + \rho^2}{\ell^2 - \rho^2}\right] \cos
\left(\frac{t}{\ell}\right).
\end{eqnarray}
The Poincar\'e coordinates can then be expressed as
\begin{equation}
z = \frac{1}{U + X}, \ \ \ \ \beta = \frac{Y}{U+X}, \ \ \ \ \gamma = \frac{T}{U+X},
\end{equation}
with the resulting line element
\begin{equation}
\label{Pm}
ds^2 = \frac{1}{z^{2}}(-d\gamma^2 + d\beta^2 + dz^2).
\end{equation}
These coordinates have two particularly useful simplifying properties.  First, the metric
(\ref{Pm}) is conformal to the Minkowski metric and, second, these coordinates
may be chosen so that the brane is located on the hypersurface $z=A$.

Note that in our various coordinate systems the moment of time symmetry can be
equivalently characterized as $t=0$, $T=0$, or $\gamma=0$.  In addition,
the $Z_2$ symmetry $\phi \rightarrow -\phi$ can be represented in Poincar\'e
coordinates as $\beta \rightarrow -\beta$.

We must now identify the boundary of the BTZ fundamental region in Poincar\'e coordinates.
To describe ${\cal F}_{\rm BTZ}$, it is useful to introduce the parameter
\begin{equation}
\label{alpha}
\alpha \equiv \frac{1}{\ell} \tanh(\pi \sqrt{M}) \ .
\end{equation}
It is easily seen \cite{Swed} that a useful choice for the boundary of the BTZ fundamental
region consists, in embedding coordinates, of the two surfaces
\begin{equation}
\label{embeddingBTZregion}
Y=\pm \ell \alpha U \ .
\end{equation}
Translating this into Poincar\'e coordinates we obtain
\begin{equation}
\label{BTZb}
\pm \beta = \frac{1}{2} \alpha \left( \ell^2 + z^2 + \beta^2 - \gamma^2 \right) \ ,
\end{equation}
with the positive sign describing the upper boundary and the negative sign
the lower one.  From now on, we shall follow only the upper boundary, as the two
are related by our $Z_2$ symmetry.

Recall that AdS$_{2+1}$ in Poincar\'e coordinates is conformally equivalent to
Minkowski space with metric $-d\gamma^2 + d\beta^2 + dz^2.$  The singularity
forms on the line where the brane ($z=A$) intersects the surface
described by (\ref{BTZb}).  In terms of the conformally rescaled space, it is
clear that this singularity propagates in a straight line (along $z=A$)
at a speed determined by (\ref{BTZb}) with $z=A$.  A short calculation
shows that this speed is
\begin{equation}
\label{singspeed}
\frac{d\beta}{d\gamma} = \frac{\alpha}{\alpha\beta +1}
\sqrt{\ell^2 + A^2 + \left(\frac{\alpha \beta + 1}{\alpha}\right)^2 - \frac{1}{\alpha^{2}}} \ .
\end{equation}
Further, the condition that the brane and the boundary of ${\cal F}_{\rm BTZ}$ be
separated at the moment of time symmetry ($\gamma=0$) yields
\begin{equation}
\ell^2 + A^2 - \frac{1}{\alpha^{2}} > 0 \ ,
\end{equation}
which, when substituted into (\ref{singspeed}), yields $d\beta /d\gamma >1$. Consequently, the singularity propagates in a straight line
at a velocity greater than that of light in the conformally rescaled metric.
In fact, the singularity lies on a surface that, in Poincar\'e coordinates
lies in the $z=A$ plane on the two hyperbolae\footnote{These hyperbolae
intersect at $\beta=0$.  To form the singular surface, one should include only
those pieces of each hyperbola which do not lie to the future of any event on the other
hyperbola.} of constant proper time
$\sqrt{\ell^2 + A^2 - \frac{1}{\alpha^{2}}}$ from the events $z=A$, $\beta =
\pm 1/\alpha$, $\gamma=0$.
It follows that no two points on the singularity are in causal
contact.

To better understand this, note that
figures~\ref{figure1} and \ref{figure2} show that, for any brane
position $A$ one can in fact choose a black hole small enough that the boundaries
of the two fundamental regions contact at $t=0$, or as close to $t=0$
as one would like.  In particular, this occurs `first' outside the horizon.
One can also show that for $A^2 > \alpha^{-2}(\alpha^{-2} -1)$
the singularity `first' develops inside the horizon if time ordering is
defined using the Poincar\'e coordinate $\gamma$.  However, since no two
points along the singularity are in causal contact, we see that this ordering
is of little relevance.  Therefore, the singularity is best thought of as arising
`simultaneously' throughout the space-time.

Our conclusion must hold in the original AdS metric as well.
It follows that the collision of black hole and brane ends not in a
black hole attached to the brane, but instead in a space-like singularity in which
much of the universe is destroyed.  The attentive reader will have
noticed something interesting about the line $\beta =0$, $z=A$,
which may be thought of as the `leading piece of the brane' in the global
coordinates of figure~\ref{figure3}.  This line reaches the singularity at the finite
time
\begin{equation}
\gamma_{\rm leading} = \sqrt{1+ A^2} \ .
\end{equation}
It may seem odd that this event is not
on the edge of the AdS space.  However, one can readily show that
the event does lie on the singularity of the BTZ black hole; in particular,
between the two horizons.  Thus, we see that the collision-induced
singularity engulfs the entire exterior space-time on the side of the black
hole that contains the brane, and that this singularity joins on to the black hole
singularity.  In fact, one could perhaps interpret this result as extending
the black hole singularity out to the edge of the space-time, destroying the
future asymptotic region and making the concept of an event horizon
meaningless on the right side of the black hole.

In any case,
one sees that our new singularity does not extend into the left exterior
region of the BTZ space-time.  The left exterior is completely
unaffected by the presence of the brane in the right exterior.
This is as it should be, since the black hole horizons should prevent
any influence from propagating from the right exterior to the left one.

The appearance of this singularity seems to demonstrate
an instability of the system that is perhaps of even more interest
than the potential failure of thermodynamics described in section
\ref{sec:raychauduri}.

\section{Discussion and Conclusions}
\label{sec:conclusions}

We have illustrated two difficulties that arise when negative tension
branes encounter black holes.  The first is a potential violation of the
generalized
second law of thermodynamics that can occur whether or not the brane is located
at an orbifold fixed plane.
Note that if the
negative energy
stored in the brane tension could be transformed into some other sort of energy,
then one could also use it to violate the second law in systems without
black holes.  For example, if the negative energy could be used to cool
a hot star, this cooling would seem to lead to a decrease in entropy.
The important feature of black holes is thus that
any form of energy that enters them directly affects the horizon area (and
thus the entropy).  The second difficulty we investigated is
a nonlinear dynamical instability that causes the entire space-time outside
the black hole to collapse
when the brane is located at an orbifold fixed plane.

We have seen that horizons of 2+1 dimensional black holes
shrink when encountering a negative tension brane if no other
matter is present.  This follows from the identical vanishing of the
shear of a null congruence in 2+1 dimensions.  One
expects similar behavior to occur in higher dimensions, unless for some
reason a significant quantity of gravitational radiation is also present
so that the shear term becomes important in the Raychaudhuri equation.

This results in a clear violation of the second law of thermodynamics if
the collision connects two equilibrium configurations.
Ideally, we would like to begin with a black hole well-separated from the brane
and end with some well-defined configuration.  We found in section
\ref{sec:space-time} that 2+1
space-times containing a moment of time symmetry with
well-separated BTZ black holes and negative tension branes
do in fact exist.  The black hole and brane then begin to fall toward one
another.  The result of this collision is the collapse of the
entire space-time outside the black hole
in a space-like singularity.  Although we have
explicitly discussed only a two parameter subfamily of the full
4-parameter family of such time symmetric initial data, continuity tells us
that the same result
follows for an open subset of such data containing our subfamily.
Similarly, the conclusion must hold for an open subset of the full
space of initial data,
which would allow for spinning black holes as well.

Let us note that the singularity forms along the brane and let us recall
that it can be thought of as forming simultaneously across the entire space-time
outside the black hole.
As a result, it is clear that at least some of the brane falls through
the horizon of the black hole before the singularity develops.  The black
hole then proceeds to `eat' the brane and reduce in size.  Where the horizon
intersects the final singularity, it is of zero size.
The area of the piece of brane that falls through the horizon
before the singularity occurs can be calculated.  Multiplying by
the brane tension gives
\begin{equation}
\mu_{\rm brane} = - 2\frac{A}{\ell} \sqrt{\frac{1}{\alpha^{2}} - \ell^2} \ ,
\end{equation}
which, roughly speaking, is the amount of negative brane mass that has
fallen though the horizon.  One can show that this is much greater than $M$.
It is natural to associate this excess with the
kinetic energy that the system acquires during the time in which the brane
and black hole are falling toward each other so that the sum of $M$, the kinetic
energy, and $\mu_{\rm brane}$ vanishes.

We would like to interpret the singularity formed in the
collision as describing a non-linear instability in the presence
of negative tension branes on $Z_2$ orbifolds.  However, this raises
certain questions.  First, we have seen that, for a fixed `brane location'
$A$, separated black holes arise only when the mass of the black hole
is sufficiently large.  As a result, one might be tempted to think
that any such instability is triggered only by large disturbances.
Recall, however, that the brane location $A$ is not physically meaningful
in and of itself.
There is in fact an AdS isometry that simply scales
the Poincar\'e coordinates, taking our brane to an arbitrary new value of $z$.
Thus, the correct statement is that black holes of any size are allowed
so long as they are sufficiently far away from the brane.

In addition, it is known that BTZ black holes can form dynamically
from the collision of matter \cite{formBTZ}.  One would therefore
naively expect that arbitrarily small BTZ black holes could in fact
be formed even close to negative tension branes.  The natural conclusion
is that the formation process can begin near the brane, but that
a singularity of the sort discussed here arises before this process is complete.
Thus, we expect non-linear
instabilities with negative tension branes in the presence of matter fields
even if no black holes are initially present.

One may ask whether the same sort of singularity that we found in 2+1
dimensions also occurs in higher dimensional cases. After all,
we have shown that no smooth
static solutions exist for black holes on positive tension branes in the 2+1 dimensional
case, while such solutions have constructed in 3+1 dimensions \cite{Emparan:2000wa}.

Let us first recall that certain black holes
\cite{Aminneborg:1996iz,Brill1996a,Brill1996b} in higher
dimensions are also asymptotically AdS and can be represented as
quotients of AdS space analogous to those that yield the BTZ black
hole in 2+1 dimensions.  Collisions of such black holes with
negative tension branes are quite similar.  The singularity again
lies in the $z=A$ plane on hyperbolae of constant proper time
$\sqrt{\ell^2 + A^2 - \frac{1}{\alpha^{2}}}$ from the events at
$z=A$, $\beta = \pm 1/\alpha$ and so forms a causally disconnected
set\footnote{One difference, however, follows from the fact that
the unusual horizon topology means that these black holes have
only a single exterior region.  In some sense, the two exteriors
of the 2+1 dimensional BTZ hole become connected.  As a result,
the fact that the left BTZ exterior was not affected by the brane
translates into the statement that the collision-induced
singularity in the higher-dimensional analogues does not engulf
the entire exterior region.  However, it does engulf the entire
brane.}. However, black holes constructed in this way have an
unusual topology for both the horizon and any associated anti-de
Sitter region \cite{Galloway:1999br}-\cite{Friedman:1993ty}.  As a result,
such black holes do not form from the
collapse of matter subject to normal boundary conditions. These
examples are therefore not sufficient to argue convincingly for a
dynamical instability in higher dimensions.  Clearly, a productive
line of investigation in 3+1 dimensions would be to perform a
stability analysis of the solution found in \cite{Emparan:2000wa}
describing a black hole on a negative tension 2+1 brane\footnote{Comparing
the entropies of black holes on and off negative tension branes (as done in
\cite{Emparan:2000wa} for positive tension branes) does suggest that the
black hole has less entropy on the brane. This is in turn suggestive of
a dynamical instability. We thank Roberto Emparan for this comment.}.

Lastly, let us recall that negative tension objects are often motivated
by considering orientifold constructions of string theory. Although orientifolds
do indeed have negative tension, it is important to note two further properties.
First, when orientifolds arise in a space-time
that is locally asymptotically
flat, the orientifolds have enough
supersymmetry (see e.g.
\cite{Hanany:2000fq}) to allow one to show that they represent
a state of minimal energy for their charge\footnote{It is in fact plausible
that orientifolds represent the absolute lowest energy state compatible with the
asymptotic metric.}.  As a result,
one strongly expects that they are dynamically stable even in the presence
of (at least uncharged) black holes.
Second, let us consider a particularly well understood orientifold known
as the O6-plane. When the coupling is strong enough, this orientifold
corresponds to a smooth classical eleven-dimensional space-time
\cite{Seiberg:1996bs,Seiberg:1996nz,Sen:1997kz} which is
${\bf R}^7$ times the Atiyah-Hitchin metric \cite{AH}.  This space-time is
Ricci-flat and satisfies the vacuum Einstein equations.
As a result, one may describe
the collision of a black hole with such an orientifold as a problem
in pure Einstein-Hilbert gravity.  Under such conditions,
the Raychaudhuri equation leads in the usual way \cite{HE}
 to the conclusion that the total horizon area increases.  As a result,
our violations of the generalized second law of thermodynamics will not
arise in this context.  As the various stringy negative tension orientifolds are
related by T-duality, one expects that the other orientifolds of string theory
also have properties such that the second law of thermodynamics
is upheld in collisions with black holes.
Since we have seen that the orbifold boundary condition itself
is not sufficient,
it would be interesting to understand in more detail just what
properties of these orientifolds enforce the second law.  For example, it
may be due the special form of the coupling of an orientifold to the
dilaton or gauge fields of string theory.

In summary, we have seen that the orbifold boundary condition of e.g.
\cite{Randall:1999ee} is not by itself sufficient to render negative tension
branes stable.  Our results indicate that both thermodynamic and non-linear
dynamical instabilities remain.
As a result, if use is to be made of negative tension branes in various models,
it is important first to show that the particular branes being used are immune
from these effects.  A conservative working hypothesis might be that
only the rather specific negative tension branes that
arise as supersymmetric orientifolds of string theory should be considered.

\acknowledgments
The authors would like to thank
Rob Myers and Lisa Randall for useful discussions.
D.M. was supported in part by
NSF grant PHY97-22362 to Syracuse University,
the Alfred P. Sloan foundation, and by funds from Syracuse
University. M.T. is supported by funds provided by Syracuse University.

\end{document}